# Image-Based Artificial Intelligence in Wound Assessment: A Systematic Review

D. M. Anisuzzaman*, Chuanbo Wang, Behrouz Rostami, Sandeep Gopalakrishnan, Jeffrey Niezgoda, and Zeyun Yu*

*Abstract*—*Efficient and effective assessment of acute and chronic wounds can help wound care teams in clinical practice to greatly improve wound diagnosis, optimize treatment plans, ease the workload and achieve health related quality of life to the patient population.* While artificial intelligence (AI) has found wide applications in health-related sciences and technology, AI-based systems remain to be developed clinically and computationally for high-quality wound care. To this end, we have carried out a systematic review of intelligent image-based data analysis and system developments for wound assessment. Specifically, we provide an extensive review of research methods on wound measurement (segmentation) and wound diagnosis (classification). We also reviewed recent work on wound assessment systems (including hardware, software, and mobile apps). More than 250 articles were retrieved from various publication databases and online resources, and 115 of them were carefully selected to cover the breadth and depth of most recent and relevant work to convey the current review to its fulfillment.*

*Index Terms*— **artificial intelligence, deep learning, wound diagnosis, wound measurement, wound systems.**

## I. INTRODUCTION

WOUND healing is a complex process and wound care is one of the major concerns in the current medical field. Clinical presentations of wound include diabetic foot ulcers, arterial and venous ulcers, pressure ulcers, surgical wounds etc. More than 451 million people in the world are affected with diabetes and this number is expected to increase to 693 million by 2045 [1]. About 15% of them are likely to develop a diabetic foot ulceration (DFU) during their lifetime [2]. Approximately 5-8% of the world population suffers from venous disease and 1% of them will likely develop venous leg ulcer (VLU) [3][4]. Pressure ulcer (PU) is responsible for a high mortality rate (29%) among elderly people [5]. Management of acute and chronic wounds are a challenge and burden to healthcare systems. A recent retrospective analysis of Medicare beneficiaries in United States demonstrated that ~8.2 million people had acute and chronic wounds at an annual cost ranging from $28.1 billion to $96.8 billion [6].

Due to the shortage of well-trained wound specialists in primary and rural healthcare settings, a large number of wound patients do not have access to specialized wound care and updated guidelines. Developments of remote telemedicine systems can greatly benefit patients in distant locations especially in rural areas with better diagnostic advices [7]. With increasing uses of artificial intelligence (AI) technologies and portable devices such as smartphones, it is now timely to develop remote and intelligent diagnosis and prognosis systems for wound care. An intelligent system can be extremely beneficial for wound care in many ways: improved precision, reduced workload and financial burden, standardized diagnosis and management and higher quality of patient care [8]. To this end, we present this systematic review that emphasizes computational methodologies behind wound measurement, and diagnosis with the latest AI technologies.

Even though there are significant literature on the research conducted in automated or semi-automated measurement, diagnosis and prognosis of wounds, very little work is available to systematically review the state-of-the-art methods in this emerging and interdisciplinary area. Existing reviews in the wound care area are mostly focused on wound-imaging techniques. Casas et al. [9], Mukherjee et al. [10], Albert [11], and Jayachandran et al. [12] reviewed the optical wound imaging technologies for assessing and monitoring wound healing. Sensors were also reviewed with imaging systems for wound healing by Dargaville et al. [13]. For wound measurement and healing, telemedicine systems were reviewed by Chanussot-Deprez et al. [14]. Additionally, there are many reviews on specific wound types (diabetic foot ulcer, pressure ulcer etc.) mainly focusing on clinical treatments [15, 16, 17, 18]. Although these reviews discussed prior work on wound statistics, characteristics, measurement, and treatment planning, no automated approaches or systems were involved.

To measure and classify wounds intelligently and make prognostic prediction depending on the wound conditions, multi-modality data are often considered as inputs, including 2D images, 3D surface shapes and textures, and texts (clinical notes, discharge summaries, and other electronic health records (EHR) etc.). Various algorithms (rule based, machine learning, and deep learning) will be discussed on feature extraction from 2D image and 3D shapes. These features are then passed as assessment parameters to wound diagnosis models for wound

.
D. M. Anisuzzaman Chuanbo Wang and Zeyun Yu are with the Computer Science Department, University of Wisconsin-Milwaukee, Milwaukee, WI 53211, USA.
Behrouz Rostami is with the Electrical Engineering Department, University of Wisconsin-Milwaukee, Milwaukee, WI 53211, USA.
Sandeep Gopalakrishnan is with the College of Nursing, University of Wisconsin-Milwaukee, Milwaukee, WI 53211, USA.
Jeffrey Niezgoda is with the AZH Wound Center, Milwaukee, WI 53221, USA.

classification, and tissue analysis. Fig. 1 shows the workflow of the reviewed wound studies.

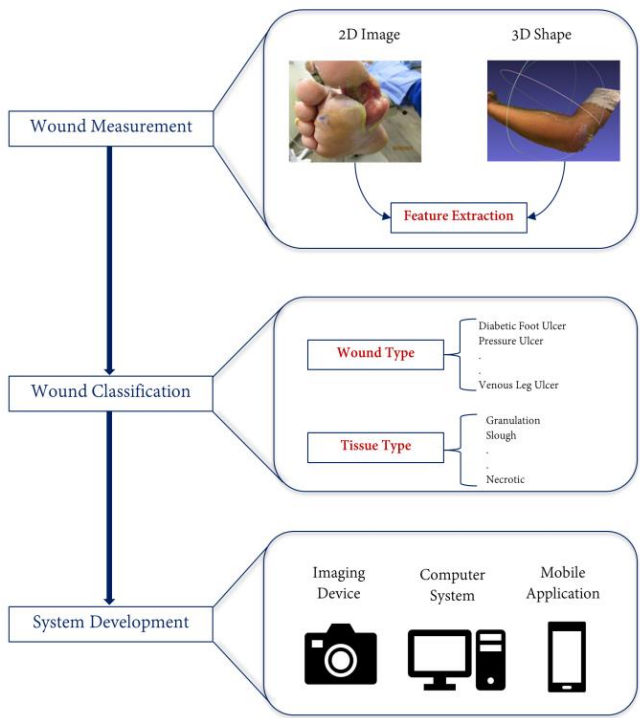

Fig. 1. Survey Workflow

We have searched popular databases (Scopus, Web of Science, PubMed, ERIC, IEEE Xplore, ScienceDirect) for papers published from January 1, 1994 to February 10, 2020 to conduct our systematic review. The keywords used for the search include wound, wound assessment, wound measurement, wound classification, wound features, wound feature extraction methods, wound predictors, wound diagnosis, developed wound care systems, pressure ulcer, diabetic foot ulcer, venous leg ulcer, burn wound, and deep learning and machine learning in wound care. In addition to research articles, some wound care web portals and applications are also reviewed. More than 250 research articles have been reviewed manually and finally 115 most related ones are selected to covey our review to its fulfillment. We have excluded the research on clinical treatment of wounds, significance of non-image wound data, and wound prognosis as the present survey is focused on the computational and intelligent data analysis of wounds and wound healing trajectory.

Wound feature extraction and wound classification involve lots of rule-based, machine learning, and deep learning algorithms. We hereby present a short summary about these methods in the biomedical engineering field.

**Rule-based algorithms** require some manually defined rules, upon which decisions can be made. For example, association rules are rule-based machine learning methods depending on if-then statements. Zhao et al. [19] presented a review of rule-based methods used for the segmentation of blood vessels. The mentioned methods for this segmentation task are: Hessian matrix, marching filtering, mathematical morphology, minimal path, active contour, and graph-based methods. Duryea et al. [20] developed a trainable rule-based algorithm for the measurement of knee joint space from radiographic images. For identifying biomarker candidates in prostate cancer, Osl et al. [21] developed a rule-based algorithm named associative voting (AV). Despite numerous rule-based algorithms available for medical (including wound) data processing, a new trend that surpasses these methods in many fields is machine learning.

**Machine learning** does not require handcrafted rules, but instead an algorithm or function is learnt from the given data. A machine learning algorithm can be supervised or unsupervised. In a supervised machine learning algorithm, the input data and the corresponding output (labels) are given; the algorithm learns itself to map the input data to the corresponding output. Some popular supervised algorithms include Bayesian network (BN), Naïve Bayes (NB), logistic regression (LR), decision tree (DT), random forest (RF), support vector machine (SVM), k-nearest neighbor (KNN), neural networks (NN), discriminant analysis (DA), single- and multi-layered perceptrons (MLP), radial basis function networks (RBF), etc. [22, 23]. On the other hand, unsupervised machine learning works with unlabeled data. Given only input data without corresponding outputs, an unsupervised algorithm learns the pattern among the data and divides them in different clusters. Some unsupervised machine learning algorithms are Markov random field, Bayesian information criterion (BIC), hierarchical clustering (GDLU, AGDL), spectral clustering, k-means, tree matching, independent component analysis (ICA), principal component analysis (PCA), decision trees etc. [24]. Giger [25] reviewed the use of machine learning in medical imaging and mentioned linear discriminant analysis, SVM, DT, RF, and NN, as some popular machine learning algorithms in this field. In the present review we regard these machine learning algorithms as traditional machine learning. By contrast, a more recent and powerful machine learning method is known as deep learning as briefly overviewed below.

**Deep learning,** as a subset of machine learning inspired by human brain, does not require any human designed rules; rather it demands a large amount of data to map the given input to specific labels (supervised learning) or clusters (unsupervised learning). Vouldimos et al. [26] carried out a survey on deep learning algorithms in the field of computer vision and they discussed some widely used algorithms: Convolutional Neural Networks, Deep Belief Networks, Deep Boltzmann Machines, and Stacked (Denoising) Autoencoders. In addition, LeNet, AlexNet, VGG 19, GoogleNet, ResNet, FCNN, RNNs, Auto-encoders and Stacked Auto-encoders, Restricted Boltzmann Machines and Deep Belief Networks, Variational Auto-encoders and Generative Adversarial Networks have been discussed as popular deep learning methods for medical image analysis [27]. Bakator et al. [28]

reviewed Convolutional Neural Networks (CNN), Restricted Boltzmann Machine (RBM), Self-Advised Support Vector Machine (SA-SVM), Convolutional Recurrent Neural Network (CRNN), Deep Belief Network (DBN), Stacked Denoising Autoencoders (SDAE), Undirected Graph Recursive Neural Networks (UGRNN), U-NET, and Class Structure-Based Deep Convolutional Neural Network (CSDCNN) as deep learning methods in the field of medical diagnosis.

The current survey will review all the three categories of approaches (rule-based, machine learning and deep learning) for wound data processing. The rest of this review follows the flow diagram of Fig. 1. Wound measurement and feature extraction methods will be discussed in Section II. Section III summarizes the work on wound classification. In Section IV, we discuss some state-of-the-art systems on wound care. Finally, we conclude the review and make some remarks on future directions.

## II. Wound Measurement

Wound measurement with some specific data sources is a necessary and important step for wound diagnosis, detecting the treatment plan, and for wound prognosis where wound healing trajectory may be predicted ahead of time. Wound measurement often involves wound feature extraction which can be divided into the following categories: feature extraction from images (2D or 3D), and feature extraction from texts (structured texts such as EHR and free texts). In this review we only consider images as the feature source.

The rest of this section is focused on the state-of-the-art works related to wound feature extraction. We shall discuss the extracted features and their sources and methods used for the feature extraction, dataset used for the validation, and limitations of these methods.

### A. Feature Extraction from 2D Images

Wound images provide the most direct and comprehensive information for physicians to diagnose a wound. There are some rule-based methods for feature extraction from 2D images by following human defined rules for specific tasks. An algorithm was developed by Pasero et al. [8] for automatic segmentation of leg ulcer using mobile phone images based on classification of pixels into two classes: "background" and "not background". The RGB and HSV color map values were used to develop this classifier. Although the algorithm has some advantages (low cost devices and fast speed), it cannot return the exact size of the wound. The Sobel edge detection method has been used to locate the affected area of ulceration [29], which can be used for detecting location, size, and shape of the wound. Wound and its location were detected from thermal images by using color-based thermal patterns [30]. Various methods (histogram equalization, Otsu thresholding and morphologic function) were used for abnormality detection at foot using a threshold obtained from the histogram. Papazoglou et al. [31] developed an algorithm based on color channels of images for determining areas of chronic wounds. The detection process requires some user intervention (manually traced wound edges). Wound dimensions can be identified by an application named WITA, developed by Damir et al. [32]. They used advanced statistical pattern recognition algorithm based on color information to extract the wound area. Fauzi et al. [33] proposed to generate a red-yellow-black-white (RYKW) probability map of an input image with a modified hue-saturation-value (HSV) model. This map is used to guide the segmentation process using either optimal thresholding or region growing. Poon et al. [34] measured the wound size by using mask images and camera calibration. They also analyzed wound color for classification, but they did not populate their model with training data. Table I gives a brief summary of wound feature extraction from 2D images, by using rule-based methods.

Traditional machine learning methods have also been used for wound feature extraction from 2D images. A cascaded two-stage classifier has been developed by Wang et al. [35] based on SVM to determine the area of diabetic foot ulcer. They used SLIC algorithm to segment the image into superpixels and then the color and texture of these superpixels are passed as the inputs of the classifiers. The first SVM classifier does a binary classification of wound and non-wound regions and the incorrectly classified instances are passed to the second SVM classifier for fine-tuning the classification. Finally, a conditional random field method was applied to refine the detected boundary. A multi-dimensional color histogram sampling technique has been used by Kolesnik et. al. [36], for segmenting wound regions. This technique generates an efficient set of feature vectors, which were passed to the SVM-classifier for reliable segmentation of wound regions. Kolesnik et al. [37] investigated the robustness of SVM classifier and added textural features on the previous SVM-based wound segmentation methods. They applied snake algorithms to adjust the initial contour produced by the SVM. Wantanajittikul et al. [38] proposed a system to segment burnt wounds from images. Cr-Transformation and Luv-Transformation were applied on the input images to remove the background and highlight the wound region. The transformed images were segmented with a pixel-wise Fuzzy C-mean Clustering (FCM) algorithm. Segmentation of wounds from color images has been done by using an SVM classifier [39]. Nine color features and one textual feature were used as the input of the SVM classifier. Song et al. [40] extracted 49 features, including area, centroid, minimum and maximum intensities, from a wound image using K-means clustering, edge detection, thresholding, and region growing in both grayscale and RGB images. These features were filtered and grouped into a feature vector that is used to train a multi-layer perceptron (MLP) and a radial basis function (RBF) neural network to identify the region of chronic wounds. Hani et al. [41] proposed to apply the independent component analysis (ICA) algorithm on pre-processed RGB images to generate Hemoglobin-based images, which were used as the input of K-means clustering to segment granulation tissues from the wound images. These segmented areas were utilized as an assessment of the early stages of ulcer healing by detecting the growth of granulation tissue on ulcer surface. Yadav et al. [42] segmented chronic wound areas by using K-means clustering and Fuzzy C-means (FCM) clustering. The dataset includes 77 images of 5 different types of wounds: PU, DFU, VLU, malignant ulcer (MU), and Pyoderma gangrenosum (PG). Their experiments showed that FCM provided better results than

4TABLE I
WOUND FEATURE(S) EXTRACTION FROM 2D IMAGES USING RULE-BASED METHODS

| Work / Research | Extracted Features | Method(s) Used | Dataset | Limitations |
|---|---|---|---|---|
| Pasero et al. [8] | wound dimension | A pixel-wise segmentation algorithm | Dataset includes 35 images from 5 patients, followed for 4 months | Unable to return the exact size of the wound. |
| Rao et al. [29] | wound size, shape and location | Sobel edge detection algorithm | Not Specified | No validation of the system was given on any dataset. |
| Kumar et al. [30] | wound area and location | Histogram equalization, Otsu thresholding and morphological method | Own data containing four sets of foot images | Very small data. Also requires thermal camera to capture thermal images whose quality depends on controllable parameters (ambient temperature, lighting, and air flow) |
| Papazoglou et al. [31] | wound area | An algorithm based on the relative color difference between the wound and its surrounding skin | Wound images from both human (99 images) and animal (56 images) studies were collected from a two-year study at Drexel University | Not suitable for the detection of nearly healed wound. |
| Damir et al. [32] | wound area | Advanced statistical pattern recognition algorithm based on color information | Not Applicable | The user has to select the wound boundary manually. |
| Fauzi et al. [33] | wound size | Red-Yellow-Black-White (RYKW) probability mapping and modified hue-saturation-value (HSV) model | Own dataset of 80 wound images, which are captured at the Ohio State University Comprehensive Wound Center in a clinical setting. | For dark skins and in some rare cases Caucasian skins leads to error. |
| Poon et al. [34] | wound size | Image masking and camera calibration | Not mentioned | Results are not validated. |

K-means clustering. Table II highlights some important factors of feature extraction from 2D images, by using traditional machine learning methods.

In recent years, deep learning had also been used for wound feature extraction. DFU segmentation has been done by using fully convolutional neural network [43]. The authors applied transfer learning and used FCN-AlexNet, FCN-32s, FCN-16s, and FCN-8s (VGG-16 based networks) as their pertained networks. They claimed to present the largest DFU dataset with annotated ground truth containing 705 foot images. The segmentation result is simply derived from the pixels classified as wounds. By testing different FCN architectures, they were able to achieve a Dice coefficient of 79.4%. Wang et al. [44] estimated wound areas by segmenting wounds with the vanilla FCN architecture. With timeseries data consisting of the estimated wound areas and corresponding images, the wound healing progress is then predicted by a function modeled by Gaussian process regression. However, the mean Dice accuracy of the segmentation is only evaluated to 64.2%. Liu et al. [45] proposed a new FCN architecture that replaced the decoder of the vanilla FCN with a skip-layer concatenation upsampled with bilinear interpolation. A pixel-wise softmax layer is appended to the end of the network to produce a probability map, which is post-processed to be the final segmentation. A dice accuracy of 91.6% is achieved on their own dataset with 950 images taken under uncontrolled lighting environment and complex background. Li et al. [46] developed a model that combined traditional methods with deep learning for wound segmentation. First, they removed the background depending on the Cr channel, then they did some preprocessing and segmented the wound using MobileNet, and finally they applied some semantic corrections (hole filling, minor noise removal, and skin detection based wound correction). Table III gives a brief summary of deep learning methods in the field of feature extraction from 2D wound images.

*B. Feature Extraction from 3D Models*

Use of 3D shapes for wound measurement provides some advantages: (1) avoidance of tedious work like molding or serum injection for wound volume measurement; (2) higher accuracy of area and volume measurements due to the 3rd dimension. Bowling et al. [7] developed a 3D wound modeling system that can help a clinician remotely measure some wound features (length, depth, volume, surface area, surface curvature, and color) through manual interaction by using a computer mouse. This study shows that a clinician can remotely assess and accurately measure a diabetic foot wound by viewing the 3D images. Wound areas, volumes, and outlines were extracted from a 3D model by Wannous et al. [47] and validated on their own dataset for robustness and reliability. Chang et al. [48] measured wound length, width, depth, surface, and volume from a 3D surface model. They constructed a 3D model of the wound with texture mapping from the RGB wound segmentation and depth view where wound size can be measured directly. Albouy et al. [49] developed a system that can construct a 3D wound model from a pair of images, from which wound inner volumes can be measured. An average plane of the wound surface was calculated by using least squares algorithm on the external boundary of the wound. The wound is then closed by this average plane, to calculate the inner volume. Treuillet et al. [50] reconstructed the textured 3D surface of pressure ulcers from two views, where no camera placement information was given. They proposed an improved version of structure from motion (SFM) algorithm which can be carried out in four stages. By closing the wound surface with a plane, wound volume can be measured with triangulation-based volume calculation from this 3D model. Pavlovčič et al. [51] developed a system to measure wound perimeters, areas, and volumes by using laser-line triangulation, color acquisition and edge detection algorithms. First, the color information of the 3D surface is converted into a 2D color image, then the wound edge is detected, followed by approximation of healthy skin and volume determination.

TABLE II
WOUND FEATURE(S) EXTRACTION FROM 2D IMAGES USING MACHINE LEARNING METHODS

| Work / Research | Extracted Features | Method(s) Used | Dataset | Limitations |
| --- | --- | --- | --- | --- |
| Wang et al. [35] | wound area, wound color and texture | SVM | Own dataset consisting of high resolution ulcer images from 15 patients over a two-year period | Images have been captured using *image-capture-box* and image capture condition, including illumination, foot position, and foot-to-camera distance have to be consistent. |
| Kolesnik et al. [36] | wound area | SVM | Trained with 6 wound images and tested on numerous images | Their method can't produce a fine wound contour compared to a clinician. |
| Kolesnik et al. [37] | wound area | SVM and Snake algorithm | Own dataset with 50 training images and 23 test images. | Very small dataset used. Maximum error was 45%. Even with both color and texture feature SVM does not guarantee a good boundary around wound. |
| Wantana-jittikul et al. [38] | wound area | Cr-Transformation, Luv-Transformation, and FCM | A dataset containing five burn images from Department of Medical Services, Ministry of Public Health, Thailand | Very small dataset has been used. |
| Kolesnik et al. [39] | wound area | SVM | Video of several wound patients' wound condition over several months. One patients' three months video gives at least 12 images | Illumination variations (brightness, lighting etc.) can affect the performance of the system. Cannot make a wound boundary as good as (manual boundary of) a clinician. |
| Song et al. [40] | wound area | MLP and RBF | A dataset containing wound images of 19 patients, collected by Drexel University from 2006 to 2008 | The robustness and reliability are yet be ensured on a bigger dataset. Determination of MLP's parameters follows no clear rules. |
| Hani et al. [41] | wound area | K-means clustering | A dataset containing wound color images, captured at Hospital Kuala Lumpur, Malaysia. | A small dataset has been used. The algorithm cannot detect all granulation regions due to specular reflections. |
| Yadav et al. [42] | wound area | K-means and Fuzzy C-means clustering | Medetec Medical Image Database | Relatively small dataset has been used and the average accuracy (72.54% to 75.23%) was not so good. |

Shah et al. [52] compared different methods for wound measurements by using rulers, 2D and 3D imaging methods. Wound volumes were calculated from 3D volumetric measurement software from the acquired 3D scanner data. Plassmann et al. [53] developed a 3D wound measurement (area and volume) system (MAVIS) by using color-coded structured light. MAVIS showed good performance on wound area and volume measurement than other systems and methods. On the next version of this system (MAVIS II) [54], a camera takes two images of a wound from slightly different viewpoints. To measure the wound dimensions including areas, volumes, and depths, special software has been used to scan both images pixel by pixel. Barone et al. [55] used a 3D optical scanner based on structured light integrated with a thermal imager for wound measurements. Kecelj-Leskovec et al. [56] developed a handheld laser-based 3D wound measuring device to compute wound perimeters, areas, and volumes. The developed device consists of a laser projector to illuminate the wound with light planes and a digital camera to take pictures from different angles. The system is non-invasive, fast, small, and easy to handle. Gardner et al. [57] examined the reliability of wound volume measurement on diabetic foot ulcer images by using VeVMD software. Though they find this software reliable, the accuracy of volume measurement has not been examined. Anghel et al. [58] have developed a 3-dimentional wound measurement (3DWM) device for wound length, width, area, depth, and volume measurements. They used an interactive graph cut algorithm for segmentation, the largest rectangle encapsulation (of the wound boundary) for length and width measurement, and a depth map (of 2D images) with depth values at pixel positions for volume measurement. Though they have found good results for wound area measurements; the accuracies of wound depth and volume were not high. Darwin et al. [59] developed a 3D device consisting of an infrared-based structure sensing camera for wound length, width, depth, perimeter, area, and volume measurements. They also used the interactive graph cut algorithm for segmentation, and depth values for volume calculation. Their system needs some manual interventions: encircling the region of interest (wound) and selecting two reference points (wound and non-wound). Table IV gives a brief summary of various methods for 3D wound measurements.

TABLE III
WOUND FEATURE(S) EXTRACTION FROM 2D IMAGES USING DEEP LEARNING METHODS

| Work / Research | Extracted Features | Method(s) Used | Dataset | Limitations |
|---|---|---|---|---|
| Goyal et al. [43] | wound area and dimension | Fully convolutional neural networks (FCNs) | Own DFU dataset with 705 images | The network is not able to detect small wounds and tend to draw smooth contours and struggled to draw irregular boundaries to perform accurate segmentation. |
| Wang et al. [44] | wound area | Vanilla FCN architecture | NYU Wound Database with over 8000 high-resolution wound images and corresponding medical records. | The mean Dice accuracy of the segmentation is only evaluate to 64.2%. |
| Liu et al. [45] | wound area | A new FCN architecture with a replacement of the decoder of the vanilla FCN. | Own dataset with 950 wound images. | Images in their dataset are annotated with watershed algorithm automatically. This means that their system is learning how watershed algorithm labels wounds in the images instead of learning from human specialists. |
| Li et al. [46] | wound area | Combination of traditional methods and DNN (MobileNet) | Own dataset containing 950 wound images, collected from internet (Medetec Medical Image Database) and hospital | Some non-skin backgrounds are misjudged as wounds. Provides too smooth segmentation results. The developed model is complex. |

Some commercial systems (tools and web portals) have also been developed for 3D wound measurements. Nixon et al. [60] referred to a 3D system named Silhouette, which can measure wound volumes, depths and perimeters. The key components of this system are SilhouetteStar™ (3D camera), SilhouetteConnect™ (3D model creating system), and SilhouetteCentral™ (internet-based record system, that passes data obtained from their data-collecting equipment to EMR). A product of eKare, named inSight 3D Measurement Module [61] can measure the area, length, width, depth, and volume of wounds from a 3D model captured by the system. This product is iPhone or iPad based, which uses infrared light for scanning. Woundvision [62] has a multimodal imaging technology named Scout, which can perform limited 3D measurements on wound images.

III. WOUND CLASSIFICATION

In this section we shall discuss the state-of-the-art wound diagnosis models based on wound images. Specifically, we will consider two research problems. The first is **wound type classification** that classifies a wound as a whole into different types (e.g., venous, diabetic, pressure etc.) or different conditions (e.g., ischemia vs. non-ischemia, infection vs. non-infection). The characterization of a wound is an essential step of wound diagnosis. The second problem considered is **wound tissue classification** that analyzes the characteristics of each pixel or a group of pixels (called super-pixels) in an image, so that pixels can be assigned to different tissue types (e.g., granulation, slough, necrotic etc.). Important diagnosis questions may be answered through the given tissue composition.

*A. Wound Type Classification*

*1. Traditional Machine Learning Methods*

Wang et al. [44] proposed a method for wound infection detection and healing status prediction using deep convolutional neural networks where the features generated by the ConvNet were used. They designed a binary classifier that fed extracted features into an SVM classifier. The output of the SVM classifier shows whether the wound is infected or not. Finally, they used a Gaussian process regression to predict the changes of wound areas for the future. Using this prediction, they can estimate the healing time of the wound. An SVM-based machine learning method has been developed by Yadav et al. [63] for burn diagnosis from burn images in the BIP_US database. They used color, texture, and depth features

TABLE IV
WOUND FEATURE(S) EXTRACTION FROM 3D MODELS

| Work / Research | Extracted Features | Method(s) Used | Dataset | Limitations |
|---|---|---|---|---|
| Bowling et al. [7] | wound length, depths, volume, surface area, surface curvature, and colors | Using the computer mouse user marked up the most important features on the wound image | A dataset containing 20 different wounds at two centers | A small dataset has been used. A manual measurement system has been applied on the 3D model. |
| Wannous et al. [47] | wound area and volume, wound outline | Mathematical calculations from 3D model which is a mesh of elementary triangles | Own dataset of chronic wounds containing several hundreds of color images | The manual viewpoint selection, some lighting conditions, and the random nature of 3D reconstruction harm the reliability of the results. |
| Chang et al. [48] | wound length, width, depth, surface, volume | 3D surface mesh building from wound segmentation and depth view | Dataset contains 133 assessment sessions from 23 subjects | A small evaluation data has been used. |
| Albouy et al. [49] | wound inner volume | 3D modeling and least squares algorithm to cover the wound surface | Dataset contains 16 pairs of images for volumetric measurement | Images are collected in controlled environment (ring flash was used and a color pattern is also used for suppressing color shifts). The color pattern is also used for volume measurement. |
| Treuillet et al. [50] | wound volume | Triangulation based volume calculation and Structure from motion (SFM) algorithm for 3D reconstruction | Only two image pairs have been used for test purpose | Very small dataset has been used. Wound outlining (required) has been done with manual tracing. |
| Pavlovčič et al. [51] | wound perimeter, area, volume | Laser-line triangulation, Canny edge detector, GrowCut and GrabCut segmentation algorithms | Four different wounds | Very small dataset used. The developed system is not ideal for clinical environment. |
| Shah et al. [52] | wound volume | 3D Reimann Sum | Six wound shaped clay plates | Result is not validated on real world (wound) data. |
| Plassmann et al. [53] | wound area, volume | Structured light approach and Cubic spline interpolation. | Own dataset of 50 patients, collected by a clinical trail | The system does not work well for very deep and very large wounds. |
| Barone et al. [55] | wound area, volume | Structured light approach | Own dataset containing wound data of seven different leg ulcer patients | Result validated on small data. The system is expensive. |
| Kecelj-Leskovec et al. [56] | wound perimeter, area, volume | Multiple-line laser triangulation; Rainbow color palette (depth measurement) | Dataset has been collected from 8 patients containing 15 VLUs | Measurement of flat wound, wound edge, and wounds with irregularities on surrounding skin is imprecise. |
| Gardner et al. [57] | wound volume, circumference, surface area, depth | VeVMD software (mathematical formula for an elliptical spheroid) | Dataset contains images of 33 patients with diabetic foot ulcer | Tracing the wound edge is hard. DFU is uniform in size and shape than other types of wounds, which may show more error in measurement (of other wound types). |
| Anghel et al. [58] | wound length, width, area, depth, volume | 3DWM device (interactive graph cuts algorithm, largest rectangle encapsulation, a depth map with depth values at pixel positions) | Dataset includes images of 45 wounds from 31 patients | The intra-class correlation coefficient (ICC) is low for depth and volume measurement. |
| Darwin et al. [59] | wound length, width, depth, perimeter, area, volume | A 3D device (interactive graph cuts algorithm, depth values representation) | Dataset includes images of 23 different kinds of wounds from 19 patients | Manual encircling of the region of interest (wound) is need for 3D transformation. |

to train their classifier to classify three types of burns: superficial dermal, deep dermal, and full-thickness burn. Abubakar et al. [64] proposed a machine learning based approach to distinguish between burn wounds and pressure ulcers. They used pre-trained deep architectures like VGG-face, ResNet101, and ResNet152 to extract the features from the images and then fed them into an SVM classifier for classifying the images into burn or pressure wound classes. The

TABLE V
WOUND TYPE CLASSIFICATION

| Work / Research | Classification | Used Features | Used Method(s) | Dataset |
|---|---|---|---|---|
| Wang et al. [44] | Binary classification of Infected and non-infected wound | extracted feature from CNN | SVM | NYU Wound Database with over 8000 high-resolution wound images and corresponding medical records |
| Yadav et al. [63] | Three types of burn: superficial dermal, deep dermal, and full-thickness burn | color, texture, and depth | SVM | Burns BIP_US Database |
| Abubakar et al. [64] | Binary classification of burn wounds and pressure ulcers | Feature extracted from VGG-face, ResNet101, and ResNet152 | SVM | A dataset containing 29 pressure and 31 burn images |
| Serrano et al. [65] | Three types of burn (superficial dermal, deep dermal, and full thickness) classification | Color and texture | Fuzzy-ARTMAP neural network | A dataset containing 38 images from 22 patients |
| Serrano et al. [66] | Two type of burns: burns that do not need grafts and burns that need grafts | color and texture (hue, chroma, kurtosis a*, skewness b*) | SVM | Burns BIP_US Database Dataset contains 20 training images and 74 testing images |
| Acha et al. [67] | Five types of burn: superficial dermal (blisters), superficial dermal (red), deep dermal, full-thickness (beige), and full-thickness (brown) | Color and texture | Fuzzy-ARTMAP neural network | A dataset containing 250 (49×49) images for feature selection and 62 (1536×1024) burn images for testing |
| Acha et al. [68] | Three burn types: superficial dermal, deep dermal, and full thickness | Mathematical features extracted from MDS physical characteristics | KNN | A dataset containing 20 burn images for feature extraction and 74 test images |
| Goyal et al. [69] | Binary classification of normal and abnormal (with diabetic ulcer) classes | Not Applicable | A novel CNN architecture named DFUNet | DFU dataset with 397 foot images (292 abnormal and 105 normal cases) |
| Aguirre et al. [70] | Binary Classification of venous or non-venous ulcer | Not Applicable | A pre-trained VGG-19 network | A dataset with 300 images with specialist annotation |
| Shenoy et al. [71] | Binary classification of each of the following nine labels: wound, infection (SSI), granulation tissue, fibrinous exudates, open wound, drainage, steri strips, staples, and sutures | Not Applicable | WoundNet (modified version of VGG-16) and Deepwound (an ensemble model) | A dataset containing 1335 wound images, collected via smartphones and internet |
| Despo et al. [72] | Four type of burn depths: superficial, partial thickness, full thickness, and undebrided | Not Applicable | Fully Convolutional Network (FCN) | BURNED containing 929 burn images |
| Alzubaidi et al. [73] | Binary classification of normal skin and abnormal skin (diabetic ulcer) | Not Applicable | A novel deep CNN (DFU_QUTNet) | A dataset containing 754 foot images |

9dataset they used in this study includes 29 pressure and 31 burn wound images obtained from internet and a hospital respectively. After augmentation they had three categories: burn, pressure, and healthy skin with 990 sample images in each class. Several experiments including binary classification (burn or pressure) and 3-class classification (burn, pressure, and healthy skin) were conducted and with the accuracy value of 99.9% as their best classification accuracy using ResNet152 for both binary and 3-class classification problems. Three types of burn depths (superficial dermal, deep dermal and full thickness) were detected by Serrano et al. [65]. First, they segmented the images by transforming them into the L*u*v* color space and applying filter smoothing, distance calculation and thresholding on the new images. From the segmented images they extracted 16 color and texture descriptors, which were fed to the fuzzy-ARTMAP neural network for three-type burn classification. In another study Serrano et al. [66] used Multidimensional Scaling Analysis (MDS) to obtain physical features from images, which are then translated to mathematical features such as chroma, outliers, hue, skewness, and kurtosis. These features are then passed to the SVM classifier to classify burns that do not need grafts (heal spontaneously) and burns that need grafts. Burns BIP_US Database was used for their experiment and they used 20 images for training and 74 images for testing. Though they achieved a high sensitivity (0.97), the accuracy (79.73%) and specificity (0.60) were poor. Acha et al. [67] classified burn images into five classes: superficial dermal (blisters), superficial dermal (red), deep dermal, full-thickness (beige), and full-thickness (brown); by using color and texture features. They selected six features (lightness, hue, SD of hue, u* chrominance, SD of v*, and skewness of lightness) by using a sequential backward selection (SBS) method and fed them to a Fuzzy-ARTMAP neural network for five-class wound classification. Acha et al. [68] used multidimensional scaling (MDS) to obtain physical features (amount of pink, texture of the color, and colorfulness) from 2D images, which were then translated to mathematical features and passed to the KNN classifier to classify three types of burn (superficial dermal, deep dermal, and full thickness).

2. *Deep Learning Methods*

Deep learning has been adopted for wound diagnosis. One of the advantages of deep learning models is that they do not require any manually crafted features as input, but rather they find the features with the given input and output. A novel CNN architecture named DFUNet was developed by Goyal et al. [69] for binary classification of two classes: healthy skin and diabetic foot ulcer skin. They applied a machine learning algorithm for feature extraction on a dataset of 397 diabetic foot ulcer images. Aguirre et al. [70], proposed a method for ulcer image classification by convolutional neural networks. In this study, a pre-trained VGG-19 network was used for classifying the ulcer images into two categories: venous or non-venous ulcer on a dataset of 300 images that were annotated by a wound specialist. Data pre-processing as well as data augmentation were conducted before the network was trained. The VGG-19 network was pre-trained using another dataset of dermoscopic images. Shenoy et al. [71] proposed a CNN-based method for binary classification of wound images. In this study they used a dataset of 1335 wound images that were collected via smartphones or downloaded from the internet. The authors considered nine different labels for the dataset and for each label two subcategories (positive and negative) was considered. They used a modified version of the VGG16 network as the classifier, named WoundNet, which was pre-trained using the ImageNet dataset. In addition, the researchers created another network, named Deepwound, which is an ensemble model that averages the results of three individual models. Finally, they provided an application for mobile phones that helps patients consult with physicians remotely about the wound healing process.

In addition to the CNN model, other deep learning methods for wound diagnosis have been discussed in literature as well. Deep learning was used by Despo et al. [72] for efficient burn classification, where they built an end-to-end deep learning model independent of any specific features of burn wounds. Also, they proposed a new wound dataset including 929 images, most of which were obtained from a medical center while the rest were collected from the internet. They defined four labels for the dataset images: superficial, superficial/deep partial thickness, full thickness, and undebrided. In the first step of their study, the authors classified the images into burn or no burn categories. The second step was related to separating the wound area from the rest of the image. In the final step, they classified the burn wound into different classes showing depth levels of each one. The authors used a modified fully convolutional network (FCN), which was built upon the VGG-16 network. Alzubaidi et al. [73] proposed a novel deep convolutional neural network, named DFU_QUTNet, for binary patch classification of normal skin versus abnormal skin including a diabetic ulcer. They also introduced a new dataset consisting of 754 foot images collected from a diabetic hospital center in Iraq. They generated 542 normal patches and 1067 patches with diabetic ulcer. In the augmentation step, they multiplied the number of training samples by 13, using transformations like flipping, rotating, and scaling. The performance of their proposed method was compared with those of other deep convolutional neural networks like GoogLeNet, VGG16, and AlexNet. Table V gives a brief summary of wound type classification.

B. *Wound Tissue Classification*

1. *Traditional Machine Learning Methods*

Wannous et al. [47] developed a multi-view strategy for tissue classification (granulation, slough, and necrosis) based on an SVM classifier, relying on a 3-D model where tissue labels were mapped, and classification results are merged. In this work, the single view classification results are merged into multi-view 3D models. A multi-class classification has been done by Veredas et al. [74] for wound-bed tissue recognition by using three machine learning approaches. In more detail, they used k-means clustering method for the segmentation part and then utilized three different classifiers including neural networks, SVM, and random forest decision trees for the classification part. They used a dataset of 113 pressure ulcer wound images. They reported high accuracy rates for the classifiers using SVM and random forest trees, but the lowest accuracy with neural networks. Veredas et al. [75] classified five types of tissue (necrotic, slough, granulation, healing, and skin) in wound regions by using a hybrid approach consisting of neural networks and Bayesian classifiers. A four-stage-cascaded binary classification was performed by neural networks followed by a Bayesian classifier. Hazem et al. [76] classified three types of wound tissue (granulation, slough, and necrosis) by using an SVM region classifier, with color and texture as input. By using the sequence selection forward method (SFS), they found 20 descriptors to be more relevant for the classifier. Mukherjee et al. [77] classified three types of wound tissue (granulation, necrotic, and slough) by using Bayesian and SVM classifiers. They considered six types of wounds (burn, DFU, MU, PG, VLU, and PU) from their database. Their experiment showed that SVM with 3rd order polynomial kernels provided the best result. Wannous et al. [78] classified four types of tissue (granulation, slough, necrosis, and healthy skin) by using color and texture features as input to the SVM region classifier. Their experiment showed that the segmentation-driven classification approach was more suitable than pixel-based approach. In their following work [79] they continued segmentation-driven classification of four tissue types by using color and texture features as input to the C-SVM classifier. Their developed tool ensures stability under various lighting conditions, viewpoints, and camera settings.

TABLE VI
WOUND TISSUE CLASSIFICATION

| Work / Research | Classification | Used Features | Used Method(s) | Dataset |
|---|---|---|---|---|
| Wannous et al. [47] | Three tissue types: granulation, slough, and necrosis | color, texture | SVM | A database of chronic wounds containing several hundreds of color images was built with expert help |
| Veredas et al. [74] | Four tissue types: necrotic, slough, healing, and granulation | color, texture, topological, and morphological descriptors | SVM, NN, and RF | Own dataset of 113 wound images, captured in non-controlled illumination conditions. |
| Veredas et al. [75] | Five tissue types: skin, healing, granulation, slough and necrosis | color, texture | Multilayer perceptrons and Bayesian classifiers | A dataset of 113 photographs containing all the tissue types |
| Wannous et al. [76] | Three types of tissue: granulation, slough, and necrosis | Color and texture | SVM | A dataset containing several hundred color pictures |
| Mukherjee et al. [77] | Three types of tissue: granulation, necrotic, and slough | Color and texture | Bayesian classifier and SVM | Medetec medical image database (74 wound images) |
| Wannous et al. [78] | Four types of tissue: granulation, slough, necrosis, and healthy skin | Color and texture | SVM region classifier and unsupervised color region segmentation | A dataset containing several hundred color pictures |
| Wannous et al. [79] | Four types of tissue: granulation, slough, necrosis, and healthy skin | Color and texture | Soft-margin SVM classifier (C-SVM) | A dataset containing several hundred color pictures |
| Zahia et al. [80] | Three types of tissue: granular, slough, and necrotic | Not Applicable | A 9-layer CNN containing 3 convolutional layers | A dataset containing 22 pressure wound images |
| Rajathi et al. [81] | Four types of tissue: granulation, slough, epithelial, and necrotic | Not Applicable | A 4-layer CNN | A dataset containing 1250 varicose ulcer images |
| Nejati et al. [82] | Seven types of tissue: necrotic, healthy granulation, slough, infected, unhealthy granulation, hyper granulation, epithelialization | An 18-dimensional feature vector produced by the DNN | AlexNet architecture and SVM patch-level classifier | Own dataset containing 350 images of chronic wounds, captured in different conditions |
| Pasero et al. [83] | Six types of tissue | Not Applicable | Self-Organizing Map (SOM) | A dataset of 92 images from 14 different patients |
| Blanco et al. [84] | Three types of wound tissue: granulation, fibrin, and necrosis | Not Applicable | QTDU (a combination of deep learning models and superpixel-driven segmentation methods) | A dataset containing 217 arterial and venous wound images |

*2. Deep Learning Methods*

Zahia et al. [80] proposed a method for tissue analysis of pressure wound images using deep CNNs, trained and tested on a dataset of 22 pressure wound images. For each image, the ground truth labels for pixels were specified by specialists. As a pre-processing task, the authors extracted the ROI from the original wound image and then removed the flashlight from the images and extract patches of 5×5 pixels from each ROI. These patches along with their ground truth labels were fed into the CNN model for training the network to predict three types of wound tissue (granular, slough, and necrotic) from an input image. The CNN used in this study is a 9-layer network with 3 convolutional layers. Based on the results reported, the slough tissue is the most challenging tissue to classify. Rajathi et al. [81] proposed a CNN-based method for tissue classification of varicose ulcer wound images. The dataset used in this study included around 1250 varicose ulcer images collected at a medical college in India and labeled by wound specialists. Their method is similar to the approach used by Zahia et al. [80], which includes three phases: data pre-processing, active contour segmentation for selecting the wound area from the skin, and CNN-based classification of the input image. In the pre-processing step, they removed the flashlight from the images using simple image processing technique like thresholding. The ROI was extracted using the active contour segmentation method. They used a 4-layer CNN for classification of wound tissue into four different types (granulation, slough, epithelial, and necrotic), based on the patches of 5×5 pixels generated from the ground truth and the segmented wound image. Nejati et al. [82] proposed a deep learning-based method to analyze chronic wound tissue and classify it into one of seven described classes, making it the first study for classifying wound tissue into more than four classes. In this study, they used a pre-trained AlexNet architecture for feature extraction from the wound tissue and then fed them into an SVM patch-level classifier for classification. The dataset they used in this study includes only 350 images. Pasero et al. [83] expanded their segmentation work [8] into tissue classification by using self-organizing map (SOM) and a bigger dataset. Blanco et al. [84] proposed a method for dermatological wound image analysis, named QTDU by a combination of deep learning models and super-pixel-driven segmentation. In this study they used a dataset with 217 arterial and venous wound images from lower limbs. The method includes three main steps. The first step is data preparation and region segmentation that includes labeling the images, constructing and augmenting the super-pixels. They provided 44,893 super-pixels, each of which was labeled with one of four classes: fibrin, granulation, necrosis, and not wound. The second step is data processing and training, in which two deep CNNs (ResNet and Inceptionv3) were trained by the data samples and 6 additional layers were added to the end of these deep architectures. The output of the final layer gives the tissue labels. They concluded that the networks pre-trained on the ImageNet dataset were trained faster in comparison to the architectures with random weights. In the final step, the pixelwise wound quantification masks were generated. In addition, the authors said that their proposed method generated better results by using the ResNet architecture in comparison with the Inceptionv3 model. The results on 179,572 super-pixels showed that the QTDU method outperformed the machine learning approaches. Table VI gives a brief summary of the wound tissue classification.

IV. SYSTEM DEVELOPMENT

In addition to the rich research in wound measurement and assessment, a number of systems involving hardware and software developments have been available for commercial and academic uses. There are some telemedical treatment systems which requires UMTS (3G) videophone with patients and home-visiting nurses on one side, and internet-based patient records and communication systems on the other (physician) side. Although some studies have been done on these systems [85, 86, 87, 88], they are not intelligent and have some limitations (e.g., extra cost and time-scheduling of home-visiting nurses), which can be overcome with intelligent systems. To this end, this section will be focused on physical devices and intelligent systems for wound assessments.

*A. Wound Imaging Devices*

Some foot imaging devices have been developed for wound assessment. Bus et al. [89] developed a physical foot imaging device for diabetic patients, but they can only take photos of planter surfaces. Hazenberg et al. [90] examined the feasibility of a photographic foot imaging device (PFID) for diabetic foot ulcer, with a small number of patients. In another study [91], Hazenberg and his team obtained foot temperature with infrared thermometer and foot images with PFID device, to demonstrate the validity and reliability of foot infection diagnosis. This device can capture the image of diabetic foot, which can be remotely assessed by a wound specialist. Nemeth et al. [92] developed a physical device with integrated mobile software to measure the wound area, but this device has some limitations such as battery problems, software bugs, uneven ambient lighting etc. The developed device can integrate a mobile within the device, which can capture the image and finally measure the area. Aldaz et al. [93] developed a hand-free chronic wound-imaging device with a combination of Google glass and android application, named SnapCap. Their system includes voice command, head tilting, and double blinking for wound image capturing. Table VII presents a short summary of the wound imaging systems.

4TABLE VII
WOUND IMAGING SYSTEMS

| System Name/ Description | Scope of the System | Tools Used (*for implementation and/or validation*) |
|---|---|---|
| Photographic foot imaging device for diabetic patients [89] | Produces three color images with different lighting conditions (diffuse, medial directed, and lateral directed) from each pair of feet. | Camera module, light sources (LEDs), mirror, glass plate, foot supports, lid, and computer |
| Photographic foot imaging device (PFID) [90] | Can take diabetic foot photos and remotely monitor the DFU healing. | Camera module, light sources, mirror, glass plate, foot supports, and a computer |
| Wound measurement device (WMD) with a custom mobile software (WoundSuite) [92] | Can calculate wound surface area. | Smartphone, four laser diods |
| SnapCap System for chronic wound photography [93] | Can capture digital image of wound without any hand-holding device. Can tag and transfer the images to a patient's Electronic Medical Record (EMR). | Google Glass (IMU, Microphone, Infrared Sensor, Eyepiece, and Camera), android smartphone |

*B. Computer-based Wound Analysis Systems*

*1. 2D wound Analysis Systems*

Some computer/desktop software tools have been developed for wound assessments based on 2D images. Among them some wound systems are based on thermal imaging because foot temperature is one of the important features in wound diagnosis. Liu et al. [94] developed a system for detecting diabetic foot complications based on simple asymmetric analysis between left and right foots from thermal images. But this method does not work with only one leg of data or if the characteristics of two legs are similar. For detecting inflammation and predicting foot ulcer from asymmetric thermal distribution analysis from two legs, an alternative technique has been developed by Kaabouch et al. [95]. This technique works for all types of feet projection. Netten at el. [96] provided some important studies of intelligent telemedicine monitoring system by developing a computer software which can detect signs of diabetic foot diseases by measuring temperature from high resolution infrared thermal images. The software still lacks some automated processing such as annotation of the boundaries of the foot.

Wound measurement is one of the most important factors of wound analysis system and there exists some computer-based 2D wound analysis systems. Wang et al. [97] developed a wound image analysis system for diabetics by using a smartphone app as the client, which takes wound pictures with an image-capturing box, and sends it to the server for analysis, and a PC as server, where the wound segmentation is processed. Rogers et al. [98] compared the standard ruler measurement (length × width) of wound areas with the measurement by the SilhouetteMobileTM (ARANZ) camera and claimed that ARANZ gave more accurate results. Pires et al. [99] developed a desktop application (compatible with Android platform) using the OpenCV library, for wound area measurement from an image captured by a smartphone camera. Oduncu et al. [100] developed computer software that can segment the wound boundary, approximate VLU position, and find the amount of slough tissue within that wound from a 2D image. They converted RGB images to HIS models as a part of their system development.

Some computer-based systems focus on broader areas than wound analysis. A wound analysis and management application named WITA has been developed by Damir et al. [32] which can classify three types of tissue (necrotic, slough, and granulation) based on color information. This application has a database that contains three types of records: patient, wound and examination. The same lighting and camera settings are the requirements for the wound image capturing which are passed as inputs to this application. Wound areas, color segmented wound tissues and healing scores have been measured by Wang et al. [101] by using three physical components: image capturing box, laptop and smartphone. They applied an augmented mean-shift-based algorithm from their previous work [102], for wound area determination. The red-yellow-black color model was used for wound tissue classification. The authors also measured healing scores by translating raw wound assessment results by using three equations. It is a real-time system and the image box reduces the illumination problem of wound image capturing. Kieser et al. [103] developed a software, named ARANZ, which provides rapid, accurate, noncontact wound measurements and data management.

*2. 3D Wound Analysis Systems*

Computer-based 3D wound analysis systems are less explored in the literature as compared to the 2D image-based systems. 3D images of diabetic foot wound have been used for remote assessment (length, depth, volume etc.) of diabetic foot ulcer. Bowling et al. [7] developed such a novel optical system, with which they were able to provide a color-calibrated, 3D image of the wound and surrounding tissues, but there were some limitations with this system such as inability to show exudation and moistness.

TABLE VIII
COMPUTER SOFTWARE DEVELOPMENT IN WOUND CARE

| System Name/Description | Scope of the System | Platform | Tools Used (*for implementation and/or validation*) |
|---|---|---|---|
| 3D imaging device [7] | Can capture the 3D image of the wound and its surrounding tissues which allows a clinician for the accurate measurements of wound length, depths, volume, surface area, surface curvature, and colors. | Computer software | A flat white plastic disc with a biocompatible glue pad on one side and a black optical pattern on the other |
| WITA [32] | Can classify three types of tissues: necrotic, slough, and granulation; and measure wound dimensions. | Computer software | C# programming language and .NET Framework |
| Multimodal Sensor System for Pressure Ulcer Wound Assessment [48] | Provides multimodal analysis and assessment including: wound segmentation, tissue classification, 3D wound size measurement (length, width, depth, surface, volume), thermal profiling (blood-flow and metabolic activities), multi-spectral analysis (oxygen saturation), and chemical sensing (skin odor measurement). | Computer System and Web Portal | A hand-held probe (Chemical sensor, light source, RGB and Depth camera, Hyperspectral camera, Thermal camera) with computer system. |
| Asymmetry Analysis for Diabetic Foot Complications [94] | Can detect diabetic foot complications from thermal images | Computer software | Thermal imaging modality with an IR camera, FLIR SC305, and a commercial digital RGB camera |
| Asymmetry Analysis-Based Overlapping for Foot Ulcer Examination [95] | Can detect inflammation and predict foot ulcer from asymmetric thermal distribution analysis | Computer software | FLIR A320 (a high-resolution thermal camera), MATLAB |
| Automated Detection of Diabetic Foot Complications [96] | Can detect signs of diabetic foot diseases by measuring temperature from infrared thermal image | Computer software | MATLAB; An experimental setup (normal camera, thermal camera, LEDs, PT1000 resistor, heating resistors etc.) |
| Wound image analysis system for diabetics [97] | Can segment the wound area. | Android mobile application (client) and computer software (server) | Smartphone, image capture foot box, GPU |
| Mobile application for wound area assessment [99] | Can measure wound area and wound contours from an image. | Computer software (compatible with Android platform) | Smartphone, OpenCV |
| Skin wound image analysis [100] | Can segment the wound boundary, approximate the position of VLU, and find the slough tissue amount. | Computer Software | A digital video camera (Panasonic NVDX 100 B) and a Fuji color scale |
| An automatic assessment system for DFU [101] | Can measure wound area, color segmented wound area, and healing score. | Computer software | Image capturing box, laptop, smart-phone, router with a peer-to-peer mode |
| The ARANZ Medical Silhouette [103] | Can measure wound surface area, depth, volume, and wound edge contour. Also collects and processes wound data, clinical history, prior treatment plans etc. | Computer software | A handheld personal digital assistant (PDA) that uses laser technology. |
| A complete 3D wound assessment tool [104] | Can perform wound measurement and tissue classification (granulation, slough, and necrosis). | Computer software | Digital camera, Macbeth color pattern |
| Derma [105] | Provides a single interface to: manage patient data (Oracle DBMS), 3D scanning, and wound measurement (wound length, volume, and surface; color segmentation) and comparison. Can measure and access the time evaluation of chronic wounds. | Computer software | A laser triangulation 3D scanner (Minolta VI910) |

Chang et al. [48] developed a multimodal sensor system combining RGB imaging, thermal imaging, 3D imaging, and chemical sensing, for pressure ulcer wound measurement and management. They developed a computer GUI and web portal for organizing wound measurement and data analysis, clinical decision making, and telehealth supporting. A 3D wound assessment tool has been developed by Wannous et al. [104] for wound measurement and three-class classification (necrosis, slough, and granulation) of wound tissues. Callieri et al. [105] developed a 3D system (Derma) for evaluating skin lesions over time. This computer software provides a single interface for patient data management, 3D scanning of the lesion region, wound measurement and comparison. This system requires a 3D scanner and hence has some limitations such as extra weight and price. Table VIII presents a short summary of the computer systems developed for both 2D- and 3D-based wound assessments.

TABLE IX
MOBILE APPLICATION DEVELOPMENT IN WOUND CARE

| System Name/ Description | Scope of the System | Platform | Tools Used (*for implementation and/or validation*) |
|---|---|---|---|
| FootSnap [106] | Can capture standardize image of the sole of the diabetic foot. | iOS mobile application | iPad |
| Mobile Wound Analyzer: MOWA [107] | Can calculate ulcer size, dimension of wound edges, and three tissue colors. Can also suggest therapeutic treatment advice per the international guideline. | iOS and Android mobile application | Smartphone |
| Mobile Application for Ulcer Detection [108] | Can detect ulcer area and ulcer location. | Android mobile application | Samsung S6 smart phone and FLIR ONE mobile thermal camera |
| Mobile based wound measurement [109] | Can segment the wound area. | Android mobile application | Samsung Galaxy Tab |
| A mHealth application for pressure ulcer documentation [110] | Provides three benefits: 1) remote consultation, 2) data organization and analysis and 3) tutorial support for caregivers who are not specialized in wound care. | Android mobile application | Android smartphones |

*C. Mobile-based Wound Analysis Systems*

Mobile devices for wound assessment are very helpful for the convenience of use, but they can be very challenging as well because of some difficulties in environmental lighting, background noise, and photo quality. In this subsection we briefly summarize some mobile applications developed for wound assessments. Some apps are made only for wound image capturing. For example, Yap et al. [106] developed a foot imaging app called "FootSnap" using iPad to standardize image capturing of the plantar surface of diabetic feet.

There are some mobile apps developed specifically for wound measurement and detection. A mobile wound analyzer application named MOWA, was reviewed by Manuel Dujovny in [107], which can calculate three tissue colors (black, yellow, and red), ulcer size, and wound edge dimensions. The edge of the wound requires a freehand drawing to generate the wound size. This app can also suggest therapeutic treatment advice per the international guidelines. Luay et al. [108] developed a smartphone app for ulcer detection. This app requires a mobile thermal camera (FLIR ONE) for image acquisition and based on temperature difference it can detect ulcerous areas on the feet. Though it can detect the ulcer and its location; this app does not work on a person who has one leg or if the characteristics of two legs are similar where both legs have foot complications.

Some other smartphone apps are designed for wound healing assessments. Hettiarachchi et al. [109] developed an android mobile application for wound segmentation and to monitor the healing of chronic wounds. Wound areas are normalized for removing effects of camera distance and angle. Friesen et al. [110] developed an android app for chronic wound care that can add a new patient, store the patient information, update wound assessments, store clinical wound history, and provide a week-by-week comparison. Table IX presents a short summary of mobile applications built for wound assessments.

V. CONCLUSION & FUTURE WORK

Computational approaches and associated software and hardware are in great demand for analyzing wound images, classifying wound tissue types, measuring wound dimensions, and monitoring changes in wound over time. This process is very time-consuming and subject to intra- and inter-rater variability when done manually. So, building an intelligent system for wound measurement, diagnosis and prognosis can save a lot of clinician time, reduce economic burden and increases patient quality of life.

In this review, we summarized the current literature on wound measurement, wound classification, and various software and hardware systems developed for wound care. Image based wound data are only considered in this review as there exist very little work on text based wound data. By reviewing the wound measurement and wound classification literature, we have summarized the current state of science, reviewed the datasets used, methodology developed, and critically listed the limitations of the study. Finally, all the developed systems and their scope, the platform on which these systems were built, and the tools used for developing these systems are briefly discussed in this review.

For better diagnosis and treatment, wound care experts often have to work with multimodal data such as 2D RGB images, 3D shape models, and clinical text notes. To the best of our knowledge, no work has utilized all modalities (images, shapes, texts) mentioned in the domain of wound care, but there exists some work that makes use of some of the multimodality data. Chang et al. [48] developed a multimodal sensor system for measuring and caring pressure ulcer wounds. Their multimodality includes data from 2D RGB images, 3D models, thermal images, and chemical sensor data. Zhang et al. [111] used multimodality for cutaneous wound tissue. For multimodal assessment of tissue oxygenation, perfusion, and inflammation characteristics, they integrated hyperspectral, laser speckle, and thermographic imaging modalities in a single-experimental setup. Bodo et al. [112] quantified soft tissue wound healing by measuring tissue oxygenation, fluid content, and blood flow by using infrared spectroscopy, bipolar bio-impedance, and Doppler ultrasound respectively. Wang et al. [113] studied the multimodality of optical microangiography (OMAG) and dual wavelength laser speckle imaging (DWLSI) system to visualize and understand microvascular, hemodynamic, and metabolic changes during cutaneous wound healing.

Looking into the future, we expect that wound assessment from a good combination of multimodality wound data is highly demanded for clinical diagnosis, prognosis and management. As data-driven technologies such as deep learning are being intensively used, building public and large datasets of wounds with proper labeling is another fundamental task in computational and clinical wound analysis. Applying the latest AI algorithms on these multimodality datasets and mobile devices will provide a remote and intelligent wound care system. The system will be greatly beneficial for healthcare staff members in wound care teams and most importantly for enhancing the quality of life of patients suffering from debilitating wounds.